\documentclass[12pt]{article}
\usepackage{bm}% bold math
\begin{document}
\title{Left-Right Model of Electroweak Interaction with one Bidoublet and one Doublet Higgs Fields}
\date{}
\maketitle
\begin{center}
\textbf{Asan Damanik\footnote{E-mail: d.asan@lycos.com}}\\
\itshape Department of Physics, Faculty of Science and Technology,\\ Sanata Dharma University,\\ Kampus III USD Paingan, Maguwoharjo, Sleman,Yogyakarta, Indonesia\\
\end{center}
\abstract{We study the predictions of the left-right model of electroweak interaction based on $SU(2)_{L}\otimes SU(2)_{R}\otimes U(1)$ gauge group by using one bidoublet and one doublet Higgs fields.  We can  reproduce the low energy phenomenology of electroweak interaction by choosing the appropriate values of vacuum expectation values of Higgs fields.  Leptons can obtain a mass via two scenarios, first via Higgs mechanism with non-zero 'hypercharge-like' $I$ in Lagrangian density mass termms and put the Yukawa couplings $G_{l}>>G_{l}^{*}$, and the second via Higgs mechanism followed by a seesaw mechanism without the requirement $G_{l}>>G_{l}^{*}$.}

\begin{flushleft}
PACs: 12.60.Cn; 12.60.Fr
\end{flushleft}

\section{Introduction}
The Glashow-Weinberg-Salam (GWS) model of electroweak interaction (standard model of electroweak interaction) which is based on $SU(2)_{L}\otimes U(1)_{Y}$ gauge symmetry group has been succesful phenomenologically. But, the GWS model is still far from a complete theory because the theory does not explain many fundamental problems such as the neutrino mass existence and the origin of parity violation in the weak interaction.  Recent experimental data on atmospheric and solar neutrinos indicate strongly that the neutrinos are massive and mixed up one another \cite{Fukuda98, Fukuda99, Fukuda01, Toshito01, Giacomelli01, Ahmad02, Ahn03}. Many theories or models have been proposed to extend the standard model.  One of the interesting models is the left-right symmetry model based on $SU(2)_{L}\otimes SU(2)_{R}\otimes U(1)$ gauge group which is proposed by Senjanovic and Mohapatra \cite{Senjanovic75}.

Using two doublets Higgs fields, Senjanovic and Mohapatra found that the Higgs potential will be minimum when we choose the asymmetric solution to the doublet Higgs fields vacuum expectation values.  Within this scheme, the presence of the spontaneous parity violation at low energy arises naturally, and the electroweak interaction based on $SU(2)_{L}\otimes U(1)_{Y}$ gauge symmetry group can be deduced from the left-right symmetry model based on $SU(2)\otimes SU(2)\otimes U(1)_{B-L}$.  Finally, the  $SU(2)_{L}\otimes U(1)_{Y}$ gauge symmetry group breaks to $U(1)_{em}$ just like GWS model.  Introducing one additional bidoublet Higgs fields in order to generate fermions masses in their model, it is possible to obtain a small neutrino mass and two massive gauge bosons $m_{W_{L}}$ and $m_{W_{R}}$ with mass $m_{W_{L}}$ $<<$ $m_{W_{R}}$ respectively.  The spontaneous breakdown of parity in a class of gauge theories have also been investigated by Senjanovic \cite{Senjanovic79}.  Left-right model of quark and lepton masses without a scalar bidoublet has been proposed which predict a double seesaw mechanism for generating tiny neutrino mass \cite{Brahmachari03}.  But, left-right model without bidoublet Higgs leads to a non-renormalizable theory.

Any viable gauge model of electroweak interactions must give an answer to the two quite different problems: (i) the breaking of symmetry from the full gauge group into electromagnetic Abelian group $U(1)_{em}$ giving a mass to the gauge bosons and then explains the known structure of weak interactions, and (ii) the mass matrices for fermions \cite{Siringo04}.  Imposing the $O(2)$ custodial symmetry with left and right Higgs fields are chosen as a doublet of $SU(2)$, the known \textit{up-down} structure of the doublet fermions masses, by insertion of \textit{ad hoc} fermion-Higgs interactions, can be obtained \cite{Siringo03}.  Another model is the approximate custodial $SU(2)_{L+R}$  global symmetry \cite{Montero06}. But, the $O(2)$ custodial symmetry leads to five dimensions operator in the mass term of Lagrangian density which lead to a non-renormalizable theory.   A theory is renormalizable if the dimension of the operator in the Lagrangian density less than or equal to 4 \cite{Mandl84, Ryder85}.  Thus, the existence of $O(2)$ custodial symmetry in electroweak theory lead to a non-renormalizable theory.  A left-right symmetry model with two Higgs bidoublet is also known to be a consistent model for both spontaneous $P$ and $CP$ violation \cite{Wu}.  The spontaneous CP phases and the flavour changing neutral currents in the left-right symmetry model also open the possibilty to obtain a large $CP$ violation in the lepton sector as well as new Higgs bosons at the electroweak scale \cite{Rodriguez}.

Motivated by the rich contents of left-right symmetry model, in this paper we evaluate the predictive power of left-right model of electroweak interaction based on $SU(2)_{L}\otimes SU(2)_{R}\otimes U(1)$ gauge symmetry with one bidoublet and one doublet Higgs fields. As far as we know, there is no paper using the left-right model of electroweak interaction by introducing one bidoublet and one doublet Higgs fields to break the symmetry.  In this model, we put both left and right fermion fields to be an $SU(2)$ doublet.  In Section 2 we introduce explicitly our model and evaluate the minimum value of Higgs potential and bounded from below in order to justify our assumption on the vacuum expectation values of the Higgs fields.  In section 3, we evaluate the predictive power of the model on the gauge bosons and the leptons masses.  In section 4, we discuss our results and its implications to the fundamental processes, and finally in section 5 we present a conclusion.

\section{The Model}

We use the left-right model based on $SU(2)_{L}\otimes SU(2)_{R}\otimes U(1)$ gauge group with the following lepton fields assignment,
\begin{eqnarray}
\psi_{L}=\bordermatrix{&\cr
&\nu_{l}\cr
&l^{-}{}\cr}_{L}, \  \psi_{R}=\bordermatrix{&\cr
&\nu_{l}\cr
&l^{-}\cr}_{R},
 \label{1}
\end{eqnarray}
where $l=e,\mu,\tau$, and the bidoublet and doublet Higgs fields as follow,
\begin{eqnarray}
\Phi=\bordermatrix{& &\cr
&w^0 &x^{-}\cr
&y^{+} &z^0\cr},\:\:\phi=\bordermatrix{&\cr
&n^{+}\cr
&k^{0}},
 \label{2}
\end{eqnarray}
which break the $SU(2)_{L}\otimes SU(2)_{R}\otimes U(1)$ down to $U(1)_{em}$.  The bidoublet Higgs field $\Phi$ transforms as $\Phi(2,2,1)$ and the doublet $\phi$ transforms as $\phi(1,2,0)$.    The vacuum expectation values of the Higgs fields are choosen as follow,
\begin{eqnarray}
\left\langle\Phi\right\rangle=\bordermatrix{& &\cr
&0 &0\cr
&0 &z\cr},\;\left\langle\phi\right\rangle=\bordermatrix{&\cr
&0\cr
&k\cr},
 \label{VEV}
\end{eqnarray}
which must satisfy the relation 
\begin{eqnarray}
Q\left\langle \Phi\right\rangle=0,
\end{eqnarray}
where $Q=T_{3L}+T_{3R}+I/2$ is the electromagnetic charge operator, such that the vacuum remains invariant under $U(1)_{em}$ gauge transformations, and $I$ is the 'hypercharge-like' which can be addresed to a new kind of quantum numbers.  For example, $I=B-L$ in left-right symmetry model based on $SO(10)$ GUT.

The general potential which consistent with renormalizability, gauge invariance, and discrete left-right symmetry, is given by,
\begin{eqnarray}
V(\Phi,\phi)=-\mu^{2}Tr(\Phi^{\dagger}\Phi)+\lambda_{1} (Tr\Phi^{\dagger}\Phi)^2-\alpha^{2}\phi^{\dagger}\phi\nonumber\\+\lambda_{2}(\phi^{\dagger}\phi)^{2}+\lambda_{3}\phi^{\dagger}\phi Tr(\Phi^{\dagger}\Phi),
 \label{3}
\end{eqnarray}
where $\mu$, $\alpha$, and $\lambda_{i}\;(i=1,2,3)$ are parameters.
After explicitly performing the calculations in order to find out the minimum value of $V(\Phi,\phi)$ which is bounded from below and take the appropriate values of the parameters $\mu,\;\alpha,\;\lambda_{1},\;\lambda_{2}$, and $\lambda_{3}$, we can have $z<<k$.

The complete Lagrangian density $L$ in our model is given by
\begin{eqnarray}
L=-\frac{1}{4}\left({W}_{L\mu\nu}{W}_{L}^{\mu\nu}+{W}_{R\mu\nu }W_{R}^{\mu\nu}\right)-\frac{1}{4}B_{\mu\nu}B^{\mu\nu} +\bar{\psi_{L}}\gamma^{\mu}\left(D_{\mu L}\right)\psi_{L}\nonumber\\ +\bar{\psi_{R}}\gamma^{\mu}\left(D_{\mu R}\right)\psi_{R} +Tr|(i\partial_{\mu}-\frac{g}{2}\tau W_{L\mu})\Phi-\frac{g}{2}\Phi\tau W_{R\mu}-g'\frac{I}{2}B_{\mu}\Phi|^2\nonumber\\+|(i\partial_{\mu}-\frac{g}{2}\tau W_{R\mu})\phi|^{2}-V(\Phi,\phi)-L_{mf},
 \label{7}
\end{eqnarray}
where $D_{\mu L,R}=i\partial_{\mu}-\frac{g}{2}\tau W_{\mu L,R}-g'\frac{I}{2}$, $I$ is the quantum number which is associated with $U(1)$ generator in left-right symmetry model, $\gamma^{\nu}$ is the usual Dirac matrices, $\tau$ is the Pauli spin matrices, $g$ is the $SU(2)$ coupling (we have take the value of $g_{L}=g_{R}=g$ due to the left-right symmetry model), $g'$ is the $U(1)$ coupling, and $L_{mf}$ is the fermion mass term.  

\section{The Gauge Bosons and Leptons masses}
\subsection{Gauge Bosons Masses}
Due to the trnsformation properties of the Higgs fields, the relevant Lagrangian density mass terms for gauge bosons ($L_{B}$) in Eq.(\ref{7}) is given by,
\begin{eqnarray}
L_{B}=Tr|-\frac{g}{2}\tau W_{L\mu}\Phi-\frac{g}{2}\Phi\tau W_{R\mu} -g'\frac{I}{2}B_{\mu}\Phi|^2+|-\frac{g}{2}\tau W_{R\mu}\phi|^{2}.
 \label{mB}
\end{eqnarray}
Substituting Eq.(\ref{VEV}) into Eq.(\ref{mB}), the Lagrangian density of gauge bosons mass terms read,
\begin{eqnarray}
L_{B}=\frac{g^{2}}{4}(z^{2}+k^{2})(W_{\mu R}^{1}+iW_{\mu R}^{2})(W_{\mu R}^{1}-iW_{\mu R}^{2})\nonumber\\+\frac{g^{2}z^{2}}{4}(W_{\mu L}^{1}+iW_{\mu L}^{2})(W_{\mu L}^{1}-iW_{\mu L}^{2})\nonumber\\+\frac{z^{2}}{4}(gW_{\mu L}^{3}-g'B_{\mu})^{2}+\frac{g^{2}(z^{2}+k^{2})}{4}(W_{\mu R}^{3})^{2}\nonumber\\+\frac{gz^{2}}{2}(gW_{\mu L}^{3}-g'B_{\mu})W_{R}^{\mu 3}.
 \label{LB}
\end{eqnarray}

To find out explicitly how Eq.(\ref{LB}) look like, we define the new fields,
\begin{eqnarray}
W_{\mu R}^{\pm}=\frac{1}{\sqrt{2}}\left(W_{\mu R}^{1}\mp iW_{\mu R}^{2}\right),\nonumber\\
W_{\mu L}^{\pm}=\frac{1}{\sqrt{2}}\left(W_{\mu L}^{1}\mp iW_{\mu L }^{2}\right),\nonumber\\Z_{\mu L}=\frac{gW_{\mu L}^{3}-g'B_{\mu}}{\sqrt{g^{2}+g'^{2}}},\nonumber\\A_{\mu L}=\frac{g'W_{\mu L}^{3}+gB_{\mu}}{\sqrt{g^{2}+g'^{2}}},\nonumber\\W_{\mu R}^{3}=X_{\mu R}.
\label{wboson}
\end{eqnarray}
Using the new fields in Eq.(\ref{wboson}), the form of Eq.(\ref{LB}) reads,
\begin{eqnarray}
L_{B}=\frac{g^{2}(z^{2}+k^{2})}{2}W_{\mu R}^{+}W_{R}^{\mu-}+\frac
{g^{2}z^{2}}{2}W_{\mu L}^{+}W_{L}^{\mu-}\nonumber\\
+\frac{(g^{2}+g'^{2})z^{2}}{4}(Z_{\mu L})^{2}+\frac{g^{2}(z^{2}+k^{2})}{4}(X_{\mu R})^{2}\nonumber\\+\frac{g\sqrt{g^{2}+g'^{2}}z^{2}}{2}Z_{\mu L}X_{R}^{\mu}.
 \label{asan}
\end{eqnarray}

From Eq. (\ref{asan}) we can see that the charged gauge bosons masses in the new fields basis which defined in Eq.(\ref{wboson}) are given by,
\begin{eqnarray}
m_{W_{R}}=g\sqrt{\frac{z^{2}+k^{2}}{2}},\nonumber\\ m_{W_{L}}=\frac{gz}{\sqrt{2}},
 \label{CB}
\end{eqnarray}
and for the neutral bosons sector we have a mixing term between $Z_{\mu L}$ and $X_{\mu R}$ fields.  The neutral boson sector can be written in matrix form:
\begin{eqnarray}
M^{2}=\bordermatrix{&Z_{\mu L} &X_{\mu R}\cr
Z_{\mu L}&\frac{(g^{2}+g'^{2})z^{2}}{4} &\frac{gz^{2}\sqrt{g^{2}+g'^{2}}}{4}\cr
X_{\mu R}&\frac{gz^{2}\sqrt{g^{2}+g'^{2}}}{4} &\frac{g^{2}(z^{2}+k^{2})}{4}\cr}.
 \label{NB}
\end{eqnarray} 
From Eq. (\ref{NB}) we then have two massive bosons with masses:
\begin{eqnarray}
m_{X}=\left[\frac{(2g^{2}+g'^{2})z^{2}+g^{2}k^{2}}{8}+\frac{\sqrt{4z^{4}g^{4}+4z^{4}g^{2}g'^{2}+g'^{4}z^{4}-2z^{2}k^{2}g'^{2}g^{2}+g^{4}k^{4}}}{8}\right]^{1/2},\nonumber\\
m_{Z_{L}}=\left[\frac{(2g^{2}+g'^{2})z^{2}+g^{2}k^{2}}{8}-\frac{\sqrt{4z^{4}g^{4}+4z^{4}g^{2}g'^{2}+g'^{4}z^{4}-2z^{2}k^{2}g'^{2}g^{2}+g^{4}k^{4}}}{8}\right]^{1/2},
 \label{damanik}
\end{eqnarray}
for the $X_{\mu R}$ and $Z_{\mu L}$ fields respectively, and the photon mass $m_{A}=0$ as required.

\subsection{Lepton masses}

In our model, the leptons can acquire a mass via the ordinary Higgs mechanism if the non-zero quantum number $I$ in the Lagrangian density mass terms are violated.  In this scheme, the leptons acquire masses only from the expectation value of the bidoublet Higgs as follow,
\begin{eqnarray}
L_{mf}=G_{l}\bar{\psi}_{L}\left\langle\Phi\right\rangle\psi_{R}+G_{l}^{*}\bar{\psi}_{L}\left\langle\Phi^{*}\right\rangle\psi_{R}+h.c.,
 \label{Lmf}
\end{eqnarray}
where $G_{l}$ and $G_{l}^{*}$ are the Yukawa couplings, and $\Phi^{*}=\tau_{2}\Phi\tau_{2}$.

If we substitute the expectation value of the bidoublet Higgs in Eq.(\ref{VEV}) into Eq.(\ref{Lmf}), then we have,
\begin{eqnarray}
L_{mf}=G_{l}z\bar{l}_{L}l_{R}+G_{l}^{*}z\bar{\nu} _{L}\nu_{R}+h.c.
 \label{fermionmass}
\end{eqnarray}

If we use the requirement that the quantum number $I$ must be zero in the Lagrangian density mass terms, then we can not use anymore Eq. (\ref{Lmf}) or Higgs mechanism to generate the leptons masses.

\section{Discussion}

Introducing one bidoublet Higgs field $\phi(2,2,1)$ and one additional doublet scalar Higgs $\phi(1,2,0)$ in the left-right symmetry model based on $SU(2)_{L}\otimes SU(2)_{R}\otimes U(1)$, we can obtain four massive gauge bosons with mass $m_{W_{L}},\:m_{W_{R}},\:m_{X}$, and $m_{Z_{L}}$, and one massless gauge boson with mass $m_{A}$ which is known as photon.  The charged gauge boson with mass $m_{W_{L}}$ and the neutral gauge boson with mass $m_{Z_{L}}$ can be associated with the known gauge bosons masses in GWS model.

To see explicitly the right charged current contribution to weak interaction at low energy, we should take the effective interaction in which the Lagrangian density is given by    
\begin{eqnarray}
L_{W}=-\frac{4G_{FL}}{\sqrt{2}}J^{\mu L}J_{\mu L}^{\dagger}-\frac{4G_{FR}}{\sqrt{2}}J^{\mu R}J_{\mu R}^{\dagger}
\end{eqnarray}
where $G_{FL}$ and $G_{FR}$ are the Fermi couplings ascociated with left and the right charged currents $J_{\mu L}$ and $J_{\mu R}$ with the Fermi couplings are given by
\begin{eqnarray}
G_{FL}=\frac{\sqrt{2}g^2}{{8m}_{W_{L}}^2},\  G_{FR}=\frac{\sqrt{2}g^2}{{8m}_{W_{R}}^2}.
 \label{Fermi}
\end{eqnarray}
Because $k>>z$, then $m_{W_{L}}<<m_{W_{R}}$.  As one can see from Eq. (\ref{Fermi}), if $m_{W_{L}}<<m_{W_{R}}$ we have $G_{F_{L}}>>G_{F_{R}}$, then the structure of electroweak interaction for charged current which is known today dominant V-A interaction can be understood as the implication of $m_{W_{L}}<<m_{W_{R}}$.  Our model also predict one new massive neutral bosons with mass $m_{X}$.

From Eq.(\ref{fermionmass}), we can see that all leptons (including neutrinos) acquiring a mass via a Higgs mechanism if the non-zero quantum number $I$ is allowed in Lagrangian density mass terms.  The masses of the charged leptons are equal to the neutral leptons (neutrinos).  Recently, as dictated by the experimental data, the neutrino mass is very small compared to its generation pair charged lepton mass.  In our model, a tiny neutrino mass can be obtained via a Higgs mechanism only if we put the Yukawa coupling $G_{l}^{*}<<G_{l}$ and the non-zero quantum number $I$ is allowed in Lagrangian density mass terms.  If we impose the requirement that the quantum number $I$ must be zero in Lagrangian density mass terms, then we can not use the Higgs mechanism anymore to generate the fermions massess.  

Anothe possibility for generating the fermions masses (charged and neutral leptons) in our model without put the Yukawa coupling $G_{l}>>G_{l}^{*}$ and allowed the non-zero $I$ is the Higgs mechanism followed by a seesaw mechanism.  In this scheme, the neutrino masses read:
\begin{eqnarray}
m_{\nu_{l}}=m_{\nu_{l}}^{D}m_{X}^{-1}m_{\nu_{l}}^{D},
 \label{seesaw}
\end{eqnarray}
where $m_{X}$ is the massive neutral gauge boson mass in Eq.(\ref{damanik}) and $m_{\nu_{l}}^{D}=G_{l}^{*}z$.  The charged leptons masses read:
\begin{eqnarray}
m_{l}=m_{l}^{D}m_{Z}^{-1}m_{l}^{D},
 \label{seesaw1}
\end{eqnarray}
where $m_{l}^{D}=G_{l}z$ and $m_{Z}$ is the neutral gauge boson mass in Eq. (\ref{damanik}).
In our model, we put the vacuum expectation value of doublet Higgs is larger than the vacuum expectation value of bidoublet Higgs in order to accommodate the maximally parity violation at low energy.  This choice for the Higggs fields vacuum expectation values do not have any conflict to the present status of the Higgs particles because its existency have not yet confirmed experimentally.

\section{Conclusion}
Introducing one bidoublet and one doublet Higgs fields in the left-right symmetry model based on $SU(2)_{L}\otimes SU(2)_{R}\otimes U(1)$ gauge group, we can explain the structure of electroweak interaction at low energy which is  dominant V-A interaction as known today.  The contribution of right charged current interaction to weak interaction is very small due to the very large $m_{W_{R}}$ compared to $m_{W_{L}}$.  Leptons can obtain a mass via two scenarios, first via Higgs mechanism with non-zero 'hypercharge-like' $I$ in Lagrangian density mass termms and put the Yukawa couplings $G_{l}>>G_{l}^{*}$, and the second via Higgs mechanism followed by a seesaw mechanism without the requirement $G_{l}>>G_{l}^{*}$.

\section*{Acknowledgments}

Part of this work was done when author as a graduate doctoral student at Graduate School of Gadjah Mada University Yogyakarta with a finacial support from BPPS Scholarship Program, Dikti Departemen Pendidikan Nasional Indonesia.  Author would like to thank Prof. J. Parry of Tsinghua University Beijing China for his very useful suggestions and corrections to the first version of this paper.

\end{document}